\documentclass[%
aps,prl,twocolumn,showpacs,superscriptaddress
]{revtex4-1}

\usepackage{graphicx}
\usepackage{dcolumn}
\usepackage{bm}
\usepackage{amssymb,amsfonts,amsmath}
\usepackage{amsmath}
\usepackage{pifont}
\usepackage{amsfonts}
\usepackage{color}

\begin{document}

\title{How Influenza's Spike Motor Works}

\author{Falko Ziebert}
\affiliation{Institute for Theoretical Physics, Heidelberg University, Philosophenweg 19, 69120 Heidelberg, Germany and \\
BioQuant, Heidelberg University, Im Neuenheimer Feld 267, 69120 Heidelberg, Germany} 
\author{Igor M. Kuli\'{c}}
\affiliation{Institut Charles Sadron UPR22-CNRS, 67034 Strasbourg, France and \\
             Institute Theory of Polymers, Leibniz-Institute of Polymer Research, D-01069 Dresden, Germany}


\hyphenation{}

\begin{abstract}
While often believed to be a passive agent that merely exploits its
host's metabolism, influenza virus has recently been shown to actively
move across glycan-coated surfaces. This form of enzymatically
driven surface motility is currently not well understood and has
been loosely linked to burnt-bridge Brownian ratchet mechanisms.
Starting from known properties of influenza's spike proteins, 
we develop a physical model that quantitatively describes the observed motility. 
It predicts a collectively emerging dynamics of spike proteins and surface bound ligands 
that combined with the virus' geometry give rise to a self-organized rolling propulsion.
We show that in contrast to a Brownian ratchet, the rotary spike drive 
is not fluctuation driven but operates optimally as a macroscopic engine in the deterministic regime. 
The mechanism also applies to relatives of influenza and to man-made analogues like DNA-monowheels and should
give  guidelines for their optimization.
\end{abstract}

\maketitle

\begin{figure}[b]
\includegraphics[width=.48\textwidth]{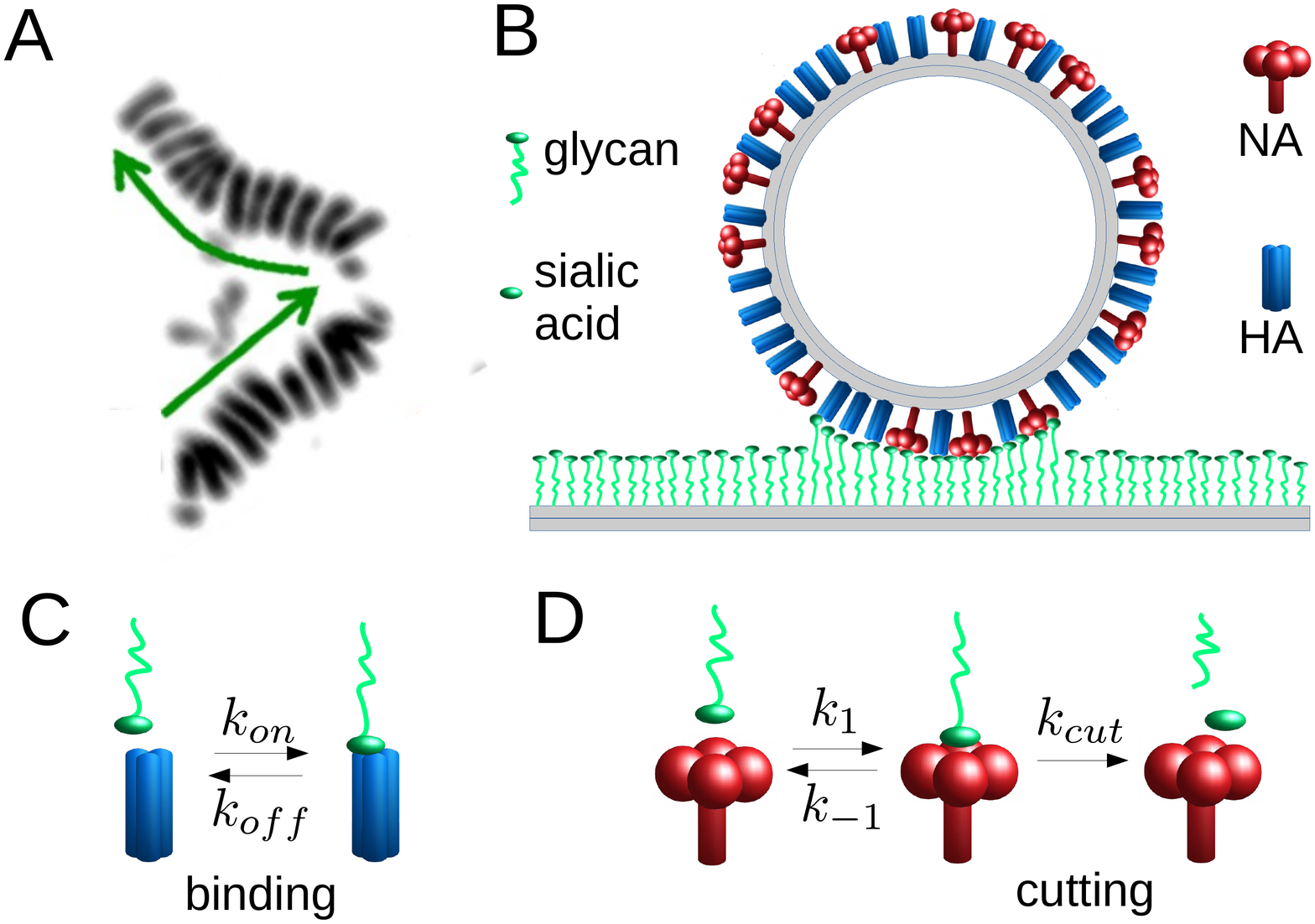}
\caption{\label{fig_IVroll}
Rolling Influenza, its surface structure and activity.
(A)  Superimposed snapshots of a rolling elongated IV C (taken from \cite{Sakai_Saito_IVC}),
the green arrows indicating the rolling direction.
(B) An IV cross-section showing its surface covered with two kinds of spike proteins: 
Hemaglutinin (HA, blue) and Neuraminidase (NA, red). The substrate (cell's surface)
is covered with glycans exposing a sialic acid residue (green, both not to scale). 
(C) HA binds to and unbinds from glycan via the sialic acid with rates $k_{on}$, $k_{off}$.
(D) NA transiently binds, with a Michaelis constant $K_M = (k_{-1}+k_{cut}) / k_1 $, 
and hydrolytically cuts the sialic residues with a rate $k_{cut}$, making the glycans inactive
for HA binding.
}
\end{figure}

One of humanity's greatest inventions is the wheel.  
While reflecting
about why nature overlooked wheeled propulsion, it caught us by surprise that the wheel 
was rolling in nature eons ago: 
the common influenza virus uses its whole capsid 
as a motorized surface rolling machine \cite{Sakai_Saito_IVA,Sakai_Saito_IVC}, see Fig.~\ref{fig_IVroll}A.
The reason for this fundamental discovery of Sakai et al.~staying almost unnoticed by
a broader audience (with few exceptions \cite{Guo_deHaan,deVries}) is possibly rooted in 
our lack of understanding of its underlying physical mechanism. 

Being such an omnipresent molecular adversary, the influenza virus (IV) and its proteins
have been extensively characterized \cite{Varghese,IV_classic_review,Gamblin_Skehel,IVAreview}. 
The two spike proteins responsible for IV A's interaction with the host membrane 
are Hemaglutinin (HA) and Neuraminidase (NA), see Fig.~\ref{fig_IVroll}B. 
While in certain influenza subtypes (like IV C) these two proteins are fused together \cite{HEFref}, 
in general they are distinct $\sim10\,$nm sized entities performing two distinct and mutually competing functions:
HA binds sialic acid residues of glyco-peptides and lipids
coating our cells (Fig.~\ref{fig_IVroll}B,C). 
NA in turn, acts antagonistically and degrades
the contacts with the glycan substrate by hydrolytically cutting the
same 
sialic acid residue that HA binds to, see Fig.~\ref{fig_IVroll}D. 
The residue, located at the very ends of branched glycans, can be either
bound by one HA or by one NA molecule, but for steric reasons not
to both at the same moment. Inhibition of HA abolishes virus binding
to glycans
(as exploited for diagnosis, e.g., in the classical hemagglutination assay \cite{Spackman})  
while NA inhibition abolishes its motility \cite{Sakai_Saito_IVC,Guo_deHaan}. 

While most textbooks depict influenza as a spheroidal
virus, its aspect ratio is in fact highly polymorphic and during infection
of human hosts the majority of the virus mass comes in filamentous
form \cite{Badham_Rossman,Dadonaite}. 
The reason for their filamentous shape with
lengths from 1-300 $\mu{\rm m}$ \cite{Badham_Rossman} 
is debated since most evolutionary arguments favor the sphere 
(e.g.~volume to surface ratio and stability \cite{Bruinsma} or uptake dynamics \cite{Frey_uptake}).
Although previously speculated \cite{Bazir} 
that elongated viruses could self-propel like man-made actively rolling fibers \cite{Baumann}, 
it was only the work of Sakai et al.~\cite{Sakai_Saito_IVC} that found the direct evidence
and suggests that the elongated form is in fact advantageous for robust directionally persistent motion.

Beyond IV and related viruses, interestingly 
DNA nano\-technology has developed synthetic rollers termed DNA-RNA monowheels \cite{DNAmw1,DNAmw2,DNArods},
that use a similar design principle -- namely linkers and their digestion.
So how does influenza, and the motif of ``bridging and cutting'' in general,
generate the force necessary for the rolling motion? Starting from basic known aspects we show here that it is not a simple burnt-bridge fluctuation driven mechanism
and that virus motion is deterministic in nature, macroscopically robust  and  in fact close-to inevitable.

\begin{figure}
\includegraphics[width=.48\textwidth]{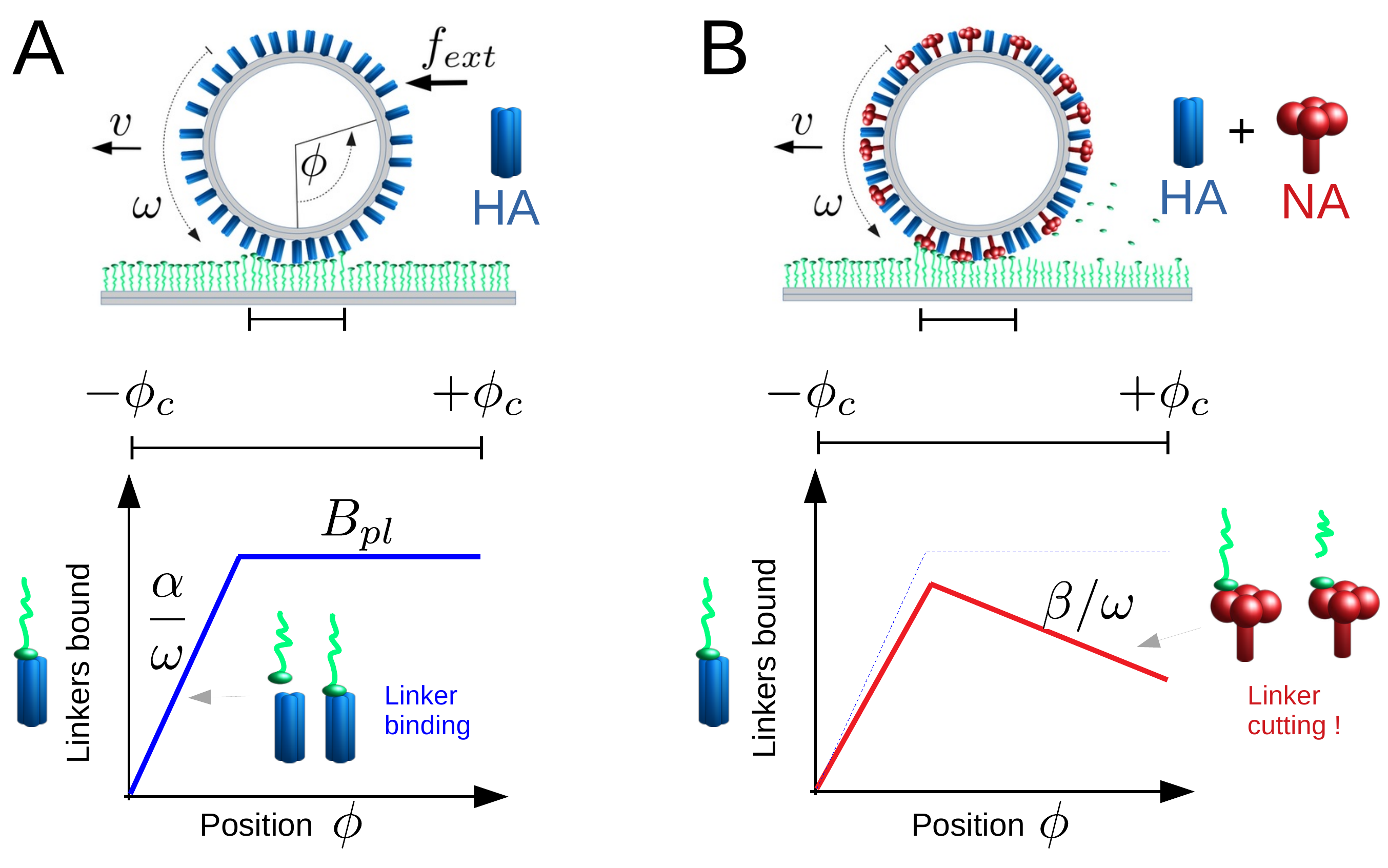}
\caption{\label{fig_mechanism}
The double-gradient mechanism of motion, maintained by the distribution of bound HA-glycans.
(A) Top: A virus rolling with constant angular frequency $\omega$ due to an externally applied force $f_{ext}$. 
Bottom: The distribution of bound HA-glycans, $B(\phi)$, within the contact interval $[-\phi_c,\phi_c]$
has two regions: a sharp increase (with slope $\alpha/\omega$) in the rolling direction 
followed by a plateau $B=B_{pl}$. 
(B) Top: self-rolling due to the enzymatic activity of NA,
cutting away the sialic acid residues. 
Bottom: In this case, the distribution $B(\phi)$ has a negative slope ($\beta/\omega$)
instead of a plateau in the second region, i.e.~at the rear.}
\end{figure}

\textit{The mechanism.} 
 Consider 
the interface where the viral capsid and the
glycan coated substrate surface meet. 
In this nanoscopic region, glycan chains at a high concentration $G_0$ 
(well in excess to spike proteins, see the estimates below), 
are constantly binding to and unbinding from the HA proteins 
and in turn elongating to a length $l$ in that process. 
Once bound (with dissociation constant $K_d$) they gain 
a free energy $k_B T \ln\left(\frac{G_{0}}{K_{d}}\right)$. On the other hand they pay a stretching energy cost 
$E_{el}\left(\phi\right)=\frac{S}{2}l^{2}\approx\frac{R^{2}S}{8}\phi^{4}$
which, due to the curvature of the capsid, depends on the angle $\phi$ measured from the virus symmetry axis 
and its radius $R$  ($\simeq 50$ nm for IV), see Fig.~\ref{fig_mechanism}A top panel. Chains bound to NA are short lived and neglected here for simplicity. 
 We consider the glycan chain as an ideal linear spring with spring constant 
$S\sim0.01$-$1\,k_{B}T/{\rm nm}^{2}$
(a typical range for polymers of few nm length). 
The balance of the two energy terms then sets 
the angular size $\phi_{c}=\left(\ln\left(\frac{G_{0}}{K_{d}}\right)\frac{8kT}{SR^{2}}\right)^{1/4}$
of the contact zone $\phi\in[-\phi_{c},\phi_{c}]$.

The stretching force $F_{el}=-\frac{\partial}{\partial l}E_{el}$ resulting from a single stretched linker gives rise to a torque 
$\propto SR^{2}\left(1-\cos\phi\right)\sin\phi\approx\frac{1}{2}SR^{2}\phi^{3}$.
The linkers have an angular density $\rho_{HA}b\left(\phi\right)$ given by the product of 
the angular density of HA spikes, $\rho_{HA}$, and 
the angular probability density of each linker being bound, $b\left(\phi\right)$,
and the total torque
acting on the capsid is the integral over all bound linkers 
\begin{equation}
m=-m_{0}\int_{-\phi_{c}}^{+\phi_{c}}b\left(\phi\right)\phi^{3}d\phi=0\,.
\label{eq:Torque m}
\end{equation}
Here $m_{0}=\frac{1}{2}SR^{2}\rho_{HA}$ is the characteristic torque
scale. At typical densities of linkers and typical angular speeds
$\omega\sim1\,{\rm s}^{-1}$ \cite{Sakai_Saito_IVC},
the linker torque dominates all other torques acting
on the virus including hydrodynamic dissipation. Therefore the  
torque balance $m=0$ holds. 

Denoting the concentration of HA spikes on the virus with $H_{0}$ 
and the initial concentration of
glycans on the (cell's) surface as $G_{0}$, 
we have to determine the evolution of the concentration
of bound HA-glycan links $B(\phi,t)$ and the 
glycan concentration $G(\phi,t)$  
both as functions of time $t$ and the angle $\phi\in[-\phi_{c},\phi_{c}]$.
In addition to this binding kinetics, the NA spike enzyme progressively
digests the glycan in its vicinity with a catalytic velocity $V_{cut}$
and a Michaelis constant $K_{M}$. Combining these effects
and assuming the virus to roll with angular velocity $\omega$ we have
\begin{eqnarray}
\partial_{t}B+\omega\partial_{\phi}B & = & k_{on}G\,\left(H_{0}-B\right)-k_{off}B\label{eq:ViroBoid B}\,\\
\partial_{t}G+\omega\partial_{\phi}G & = & -k_{on}G\,\left(H_{0}-B\right)
+k_{off}B-\frac{V_{cut}G}{K_{\mathrm{M}}+G}\,.\,\,\quad\label{eq:ViroBoid G}
\end{eqnarray}
Here the terms $\propto\omega$ on the l.h.s.~represent the advection
of concentrations in the virus fixed-frame due to its rotation. 
The first terms on the r.h.s.~are the on/off-kinetics of glycan binding, 
with the kinetic constants satisfying $\frac{k_{off}}{k_{on}}=K_{d}$. 
 Although the off-rate is stretching force- \cite{EvansRitchie}
and hence angle-dependent, we neglect this effect
assuming a small size of the contact interval $ \phi\ll1$ where
the elastic energy dependence $E_{el}\propto \phi^4$ is weak.

Finally, the last term in Eq.~(\ref{eq:ViroBoid G}) represents the
Michaelis-Menten-like degradation of the free glycans by NA, 
with a velocity $V_{cut}=k_{cut}N$ set by the cutting rate $k_{cut}$, see Fig.~\ref{fig_IVroll}D, 
and the enzyme concentration $N$.  
These two equations, together with $b\left(\phi\right)=B/H_{0}$ satisfying
the torque balance, 
Eq.~(\ref{eq:Torque m}), 
completely 
determine the dynamics and the question is now, whether
the enzymatic activity can sustain solutions with non-zero $\omega$.

\textit{Passive frictional torque.} In a fist step, it is instructive to consider
the passive case, i.e.~in absence of catalytic activity ($V_{cut}=0$),
and to assume that the virus is forced by a weak external force/torque to roll with a steady-state 
angular velocity $\omega$ \cite{Note Contact Zone}. 
Related situations have been investigated when modeling 
cells rolling in external shear flow \cite{Hammer04,KornSChwarzPRE}
and contraction/sliding motion induced by stochastic linkers \cite{Walcott,Sens_stickslip}. 
In the steady state, Eqs.~(\ref{eq:ViroBoid B}, \ref{eq:ViroBoid G}) 
imply the conservation law $\left(B+G\right)'=0$ or $G=G_{0}-B$ (for a homogeneous $G_{0}$),
allowing us to reduce the problem to 
\begin{equation}
\omega\,B'=k_{on}\left(G_{0}-B\right)\,\left(H_{0}-B\right)-k_{off}B. 
\end{equation}
For the initial condition $B(-\phi_{0})=0$
(rolling to the left) the exact solution is given by 
\begin{equation}
B\left(\phi\right)=\frac{C_{0}-C_{1}}{2}
-\frac{C_{1}}{\frac{C_{0}+C_{1}}{C_{0}-C_{1}}e^{\frac{C_{1}k_{on}}{\omega}\left(\phi+\phi_{0}\right)}-1}
\label{eq:B-equil}
\end{equation}
with constants $C_{0}=H_{0}+G_{0}+K_{d}$, $C_{1}=\sqrt{C_{0}^{2}-4H_{0}G_{0}}$. 
This profile, cf.~the sketch in the  bottom panel of Fig.~\ref{fig_mechanism}A,
implies an increase of the bound HA-glycan links leveling to a plateau 
$B_{pl}=\frac{C_{0}-C_{1}}{2}$ \cite{comment_static}. 
For simplicity,
we approximate the exact profile by two lines: 
first, in the region of its rapid increase,
$B$ is approximated by the slope at the front: 
for $\phi\in[-\phi_c,\phi_{pl}]$,  $B\left(\phi\right)= \frac{\alpha}{\omega}(\phi+\phi_c)$ 
with $\phi_{pl}$ the angle where in this approximation the plateau is reached and the linker binding velocity
\begin{equation}
\alpha=k_{on}H_{0}G_{0}.
\end{equation}  
Note that the faster the rolling, the shallower the spatial gradient becomes,
since the build-up of HA-glycan links needs time.
And second, for $\phi\in[\phi_{pl},\phi_c]$ we approximate $B$ by its plateau value, 
i.e.~$B\left(\phi\right)= \frac{\alpha}{\omega}(\phi_{pl}+\phi_c)=B_{pl}$.

With this slope-plateau approximation, for slow rotation (implying a steep increase of $B$ to the plateau), 
we can evaluate Eq.~(\ref{eq:Torque m}) to get the torque-velocity relation for passive rolling
\begin{equation}
m_{diss}\left(\omega\right)=-\xi_{diss}\omega\,\,,\,\,\,\,
\xi_{diss}=\frac{m_0}{H_{0}}\frac{B_{pl}^{2}}{\alpha}\frac{\phi_{c}^{3}}{2}.
\end{equation}
Hence this is a frictional torque, acting against the motion
and linear in $\omega$. The friction constant, $\xi_{diss}$,
is determined by both the slope and the plateau of the distribution of $B$
as well as the size of the contact interval, which themselves contain 
all system parameters.

\textit{Enzyme activity induces active torque.}
The effect of the NA activity can be captured perturbatively
and leads to an additional contribution to the torque. 
If the enzyme activity is slow compared to the  binding kinetics,
$\epsilon
=V_{cut}/\alpha$ is a small parameter. 
Expanding $B=B^{(0)}+\epsilon\,B^{(1)}$ and $G=G^{(0)}+\epsilon\,G^{(1)}$ in powers of $\epsilon$
yields the leading order correction $B^{(1)}=-\alpha f\,\frac{\phi+\phi_c}{\omega}$ with 
\begin{equation}
 f= \frac{H_0-B_{pl}}{K_d+G_{pl}} \frac{  G_{pl}}{K_M+G_{pl}}
\end{equation}
a dimensionless ratio of 
all concentrations/kinetic constants and 
$G_{pl}=G_0-B_{pl}$ the plateau of the glycan distribution (in the passive case).
The enzymatic activity hence leads to a {\it negative slope} $\beta=V_{cut} f$ instead of the plateau,
cf.~Fig.~\ref{fig_mechanism}B, and
insertion into Eq.~(\ref{eq:Torque m}) yields the {\it active torque}
in ``two-slope'' approximation:
\begin{equation}\label{mactfact}
 m_{act}=\frac{p_{act}}{\omega}\,,\,\,
 p_{act}=\frac{m_0}{H_0}f\,
\frac{2\phi_{c}^{5}}{5}  V_{cut}\,,
\end{equation}
where $p_{act}$ is the power injected by NA operation.
The active torque is positive (since $B_{pl}<H_0$), it is proportional to $V_{cut}$ and has a  $1/\omega$ dependence,
unlike the passive one which is linear in $\omega$. 

Taken together, the passive and active torques yield the
torque balance 
$m_{diss}+m_{act} =0=
-\xi_{diss}\omega+p_{act}/\omega$, 
implying a pitchfork bifurcation for the
steady-state rolling velocity
\begin{equation}\label{om_virus}
\omega=\pm\sqrt{\frac{p_{act}}{\xi_{diss}}}\propto\phi_c\sqrt{f}\,\frac{\sqrt{\alpha V_{cut}}}{B_{pl}}\,.
\end{equation}
Note that the torque scale, $m_0=\frac{1}{2}SR^{2}\rho_{HA}$, 
cancels out, but the parameters $S$, $R$ are still present due to the dependence in $\phi_c$.

We can compare to the experiments \cite{Sakai_Saito_IVC} by 
inserting typical parameters for IVs:
using $R=50$ nm and $S\simeq0.1\,k_BT/{\rm nm}^2$ 
implies $\phi_c\simeq0.5$.
Typical concentrations are $G_0=10$ mM, $H_0=1$-$5\,$(we use 2) mM  \cite{discussGH}. 
The HA on-off kinetics has been characterized \cite{Sauter1,Sauter2} yielding
$K_d=1$-$5\,(2)$ mM, $k_{off}=10^{-1}$-$1\,(1)\,{\rm s}^{-1}$, $k_{on}=0.01$-$1\,(0.5)\,{\rm mM}^{-1}{\rm s}^{-1}$
and NA's enzymatic activity \cite{Adams} to yield $K_M=14.3$ mM, $k_{cat}\simeq15\,{\rm s}^{-1}$,
implying with a typical NA concentration of $N=1$ mM a 
$V_{cut}=k_{cat}N=15\,{\rm mM}{\rm s}^{-1}$.
Using these values in Eq.~(\ref{om_virus}) we get $\omega\simeq0.4\,{\rm s}^{-1}$, which
compares well to Ref.~\cite{Sakai_Saito_IVC} where virus speeds of $v=10-30\,{\rm nm}/{\rm s}$
were reported, corresponding to angular velocities $\omega=\frac{v}{R}\simeq0.2-0.6\,{\rm s}^{-1}$.

{\it Numerical study including stochasticity.}
To scrutinize the robustness of the mechanism, so far described on the continuum
level via concentrations,
we implemented the stochastic reaction kinetics using the Gillespie algorithm \cite{Gillespie}.
For the latter, the virus cross-section was assumed to present a number of $n_{vir}$ 
discrete binding sites per (angular) contact interval $[-\phi_{c},\phi_{c}]$. 
Larger $n_{vir}$ correspond 
to a more elongated virus with more linkers per angle,
with $n_{vir}\rightarrow\infty$ being the deterministic limit. 
The virus position 
is updated in each step 
in accordance to the vanishing torque condition.

\begin{figure}[t]
\includegraphics[width=.45\textwidth]{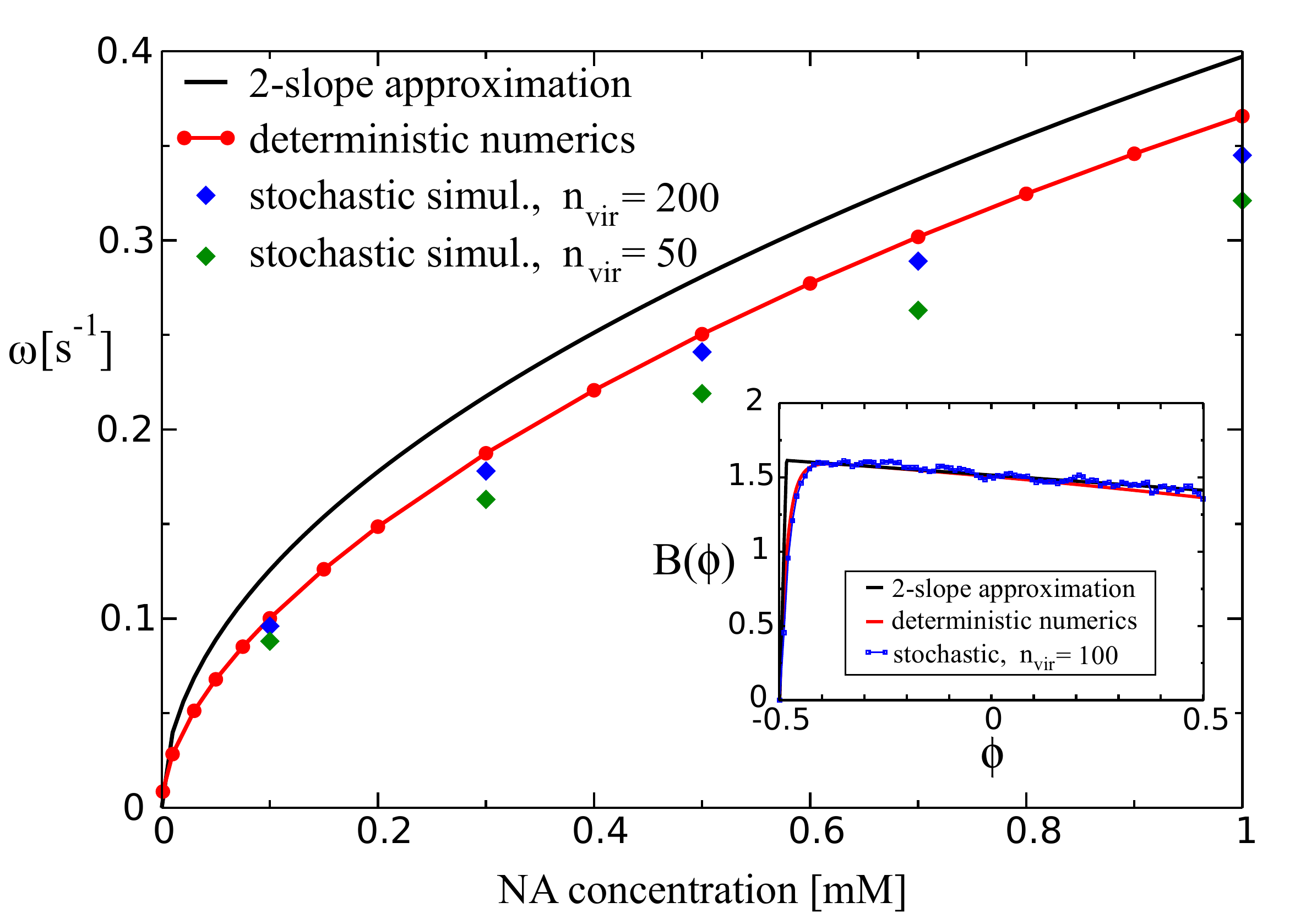}
\caption{\label{BifurcationAndProfile}
Rolling velocity and bound linker profile in the continuum model vs.~the stochastic implementation.
Shown is the rolling frequency $\omega$ as a function of enzymatic activity, 
i.e.~concentration of NA, comparing the approximate theory (black), a numerical solution of the  continuum model (red) 
and the stochastic implementation (symbols) for different numbers of discrete binding sites, $n_{vir}$.
The inset shows the profiles $B(\phi)$ for stationary rolling in the three cases.
Here $N=0.1$ mM and the stochastic profile was averaged over 50000 Gillespie moves.
}
\end{figure}

Fig.~\ref{BifurcationAndProfile} compares the approximate two-slope theory (black curve)
to a numerical solution of the continuum model Eqs.~(\ref{eq:ViroBoid B}, \ref{eq:ViroBoid G}) (red) 
and stochastic simulations (symbols).
Shown is the angular frequency $\omega$ as a function of the NA concentration.
The inset displays the profiles $B(\phi)$ for stationary rolling in the three cases.
While in the stochastic implementation the virus inverts its rolling direction occasionally,
from Fig.~\ref{BifurcationAndProfile} it is evident that the mechanism is robust against finite 
number of binding site effects. 
It nevertheless works best (i.e.~rolling is fastest) for the ``macroscopic'' i.e.~continuum case,
in contrast to the classical burnt-bridge mechanism, as discussed below. 

\begin{figure}[t]
\includegraphics[width=.45\textwidth]{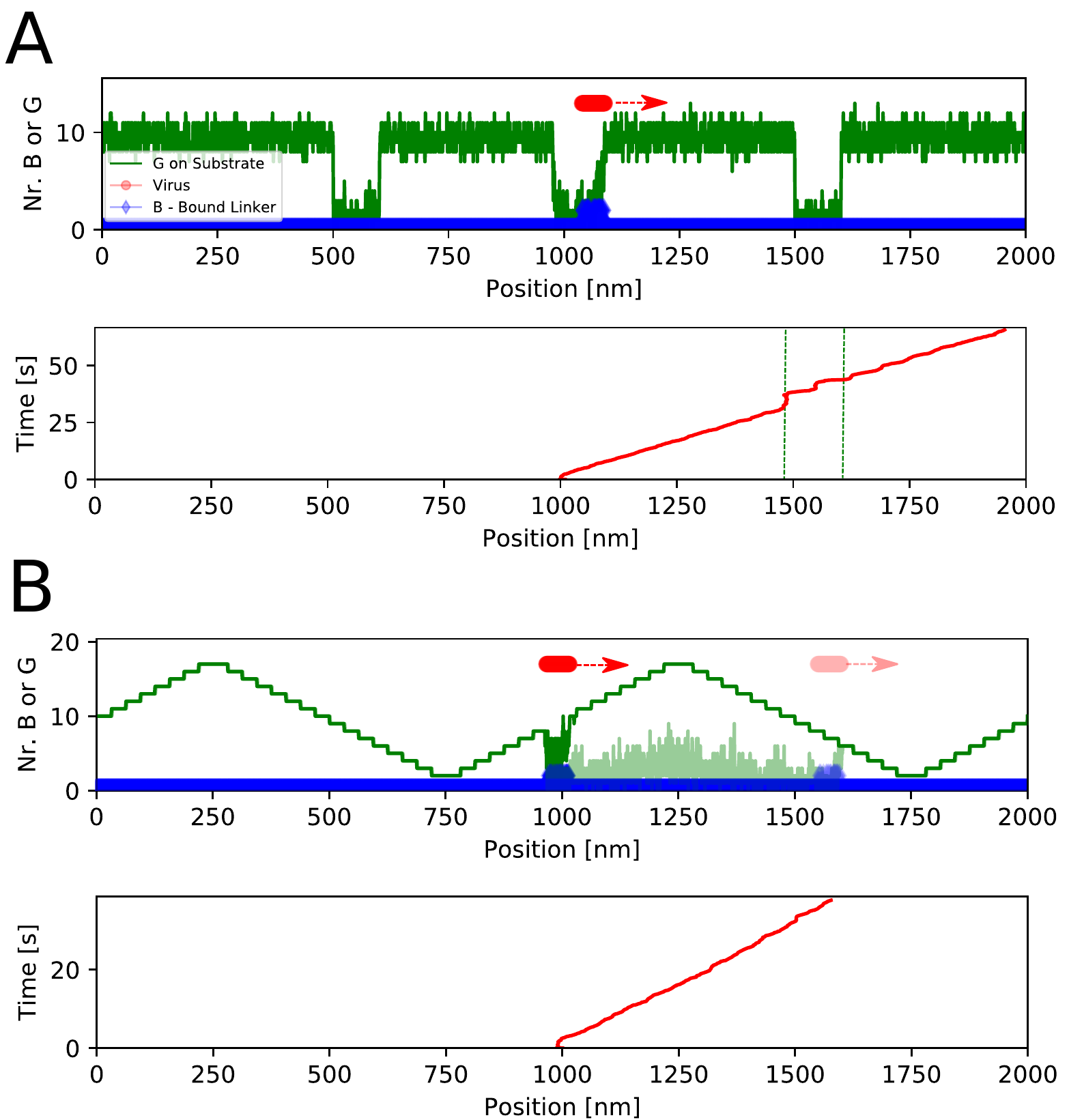}
\caption{\label{GillespieSimulations}
Stochastic simulations for modulated glycan distributions.
(A) The upper panel shows a snapshot of the bound linker (in blue) and the glycan (green) distributions
and the lower panel the trajectory. The surface-bound glycans were locally depleted (marked also 
in the trajectory by the green vertical lines), but the rolling motion persists. 
(B) Two superimposed snapshots of a virus rolling on a surface presenting uphill and downhill glycan gradients
and the corresponding trajectory below,
again showing the robustness of the rolling mechanism.
}
\end{figure}

Fig.~\ref{GillespieSimulations} shows stochastic simulations
highlighting the robustness of the mechanism against perturbations 
in the glycan distribution on the surface: the upper panel in A shows a snapshot of a virus rolling
over a surface displaying small, almost glycan-depleted regions and the lower panel the trajectory.
Although the virus is slowed down in the depleted zones (cf.~the green lines in the trajectory panel),
its motion persists.
Fig.~\ref{GillespieSimulations}B displays two superimposed snapshots of a virus moving on a 
modulated glycan distribution, the lower panel again showing the trajectory.
This proves that the mechanism allows rolling on both uphill and downhill glycan gradients.

The generic double-gradient mechanism employed by the IV is 
the interplay of binding and digestion of linkers, the
enzymatic reaction transforming the plateau of the linker distribution 
(cf.~Fig.~\ref{fig_mechanism}A) into a negative slope (cf.~Fig.~\ref{fig_mechanism}B).
One can imagine other reaction pathways to result into a similar generic 
``polarization'' with two different slopes for the bound linker distribution,
and in fact this was recently realized \cite{DNAmw1}:

{\it A synthetic relative: the DNA-RNA monowheel}. Interestingly, without knowing about the mechanism
of IV rolling, a synthetic variant of the generic mechanism was recently implemented
using DNA nanotechnology \cite{DNAmw1,DNAmw2}.
There, the surface of silica particles was covered with DNA sequences 
that form heteroduplexes (of size $\sim15$ base pairs)
with complementary RNA strands coated on a surface. 
Motion of the particles (termed {\it monowheels} in \cite{DNAmw1})
was initiated by adding
RNase H, which selectively hydrolyses the hybridized RNA (the bound linkers, $B$ in our notation), 
but not single-stranded RNAs (i.e.~free linkers, $G$).

This variant can be easily cast into our theoretical framework:
as the enzyme RNase H does not destroy the free linkers but instead the \textit{bound} ones,
the enzymatic term ($\propto V_{cut}$) is absent in Eq.~(\ref{eq:ViroBoid G}),
but its analogue $-\frac{V_{cut}B}{K_{\mathrm{M}}+B}$ has to be added to Eq.~(\ref{eq:ViroBoid B}).
This only slightly modified model can be analyzed along the same lines
leading to an active torque 
as in Eq.~(\ref{mactfact}) but 
with
\begin{equation}
 f=\frac{B_{pl}}{K_M+B_{pl}}.
\end{equation}
We can again insert numbers, namely $R=2.5\,\mu$m \cite{DNAmw1}, 
$S\simeq 0.1\,k_BT/{\rm nm}^2$ implying $\phi_c\simeq0.1$, and for the binding parameters 
$G_0=1$-$2$ mM, $H_0=2$-$4$ mM (estimated from \cite{DNAmw1}), 
$K_d=10^{-5}$ mM \cite{Bielek_Holyst,DNAmw2},
$k_{off}=10^{-5}\,{\rm s}^{-1}$ \cite{Strunz,Bielek_Holyst}, 
implying $k_{on}=1\,{\rm mM}^{-1}{\rm s}^{-1}$.
RNase H kinetics is known 
\cite{Fang}: $K_M=0.3\cdot10^{-3}$ mM, $k_{cat}\simeq1\,{\rm s}^{-1}$ 
and its typical concentration used in \cite{DNAmw1} is  $R_H=1.44\cdot10^{-4}$ mM,
implying $V_{cut}=k_{cat}R_H=1.4\cdot10^{-4}\,{\rm mM}{\rm s}^{-1}$.

For these values we get $B_{pl}\gg K_M$
hence
$f\simeq1$ and, as the passive torque has the same dependence as before,
\begin{equation}\label{omegaDNA}
\omega=\pm\frac{2}{\sqrt{5}}\frac{ \sqrt{\alpha V_{cut}} }
 { B_{pl} } \phi_c\,.
\end{equation}
Inserting numbers yields $\omega\simeq10^{-2}\,{\rm s}^{-1}$
which once again fits well to the experimentally
observed  velocity of $30\,{\rm nm}/{\rm s}$ \cite{DNAmw1} implying $\omega\simeq10^{-2}\,{\rm s}^{-1}$.

{\it Discussion.} 
The double-gradient mechanism described here is very robust and gives rise to large 
propulsion speeds, $\omega \propto \sqrt{V_{cut}}$, even for weak enzymatic activity. 
Importantly, it is not a simple
bridge-burning as recently hypothesized for both IV and the DNA-wheel \cite{Sakai_Saito_IVC,DNAmw2}.  
Although 
bridge-burning can be operative for certain IV A strains 
that show phase separation of NA and HA spikes \cite{Fletcher}, these lack rolling and move much slower 
than reported in \cite{Sakai_Saito_IVC,Sakai_Saito_IVA}. 
In burnt-bridge Brownian models \cite{Blumen,Krapivsky} the random walker destroys the ``bridges'' it walks on, 
and only the prohibited back-stepping leads to the directed motion 
implying self-avoiding paths.
Such models, like other members of the class of  Brownian motors \cite{Reimann, Ajdari}, 
are inherently
fluctuation driven and increasing the angular density of linkers implies a slow-down of the motion.    
In contrast, the mechanism discussed here relies on the self-organized, ``internal'' polarization of the linker
distribution within the  contact zone and works optimally in the macroscopic regime. 
It is robust (cf.~Fig.~\ref{GillespieSimulations})
but not self-avoiding/unidirectional, since the roller can run in 
reverse direction even if its trail 
(the glycan or RNA distribution) is substantially depleted behind.
It responds to existing surface gradients,
but much less than a burnt-bridge walker, giving the virus -- controlling its contact zone -- 
a 
higher motile autonomy in the evolutionary race with 
its host, controlling the rest of the substrate.      
How the ``delicate balance'' \cite{deVries} of NA vs.~HA and its adaptation 
orchestrates the mechanism in detail should be explored more in the future.

In conclusion, it seems that with influenza we are facing an
underestimated, 
smart adversary that in contrast to classical virology dogmas displays a 
``metabolism'' at its interface, providing it with a {\it motile organelle}
(engulfing its whole body) that emerges from geometry and the self-organization
of its spike proteins.
The mechanism should also apply to many relatives of influenza bearing enzymatic spike proteins, 
including  toro-virus and some beta-corona-viruses \cite{HEFinCoronaToroV}.
However, we can flip the coin and turn the adversary into an
ally, by learning from its workings: The triplet of a binding
molecule, a cutting enzyme and the spherical or cylindrical geometry
should be abundant
and was in fact already used to propel DNA-coated beads \cite{DNAmw1,DNAmw2},
which could now be optimized using the understanding of the mechanism.

\begin{acknowledgments}
{\it Acknowledgements.} 
We thank Jens-Uwe Sommer, Ulrich Schwarz and Felix Frey for discussions.
IMK thanks the Leibniz-Institute IPF Dresden for hospitality.  
\end{acknowledgments}


\begin{thebibliography}{99}
\bibitem{Note Contact Zone}
Note that we apply a quasi-static approximation here,
neglecting that the contact zone can become asymmetric at larger frequencies \cite{Hammer04}, 
leading to additional dissipation at the rear end where chains do not have sufficient time to detach.
\bibitem{comment_static}
Note that the bound linker-profile
in the static case, $\omega=0$, 
is a 
rectangular function with $B_{pl}$ in the contact interval, 
due to the neglection of angle-dependence of the off-rate.
\bibitem{discussGH}
The volume concentrations $G_0$, $H_0$ and $N$ (in ${\rm M}={\rm mol}/{\rm l}$) can be estimated from 
typical volumes per molecule $V=A\delta$, with $A\sim 100-500\,{\rm nm}^2$ (for glycans and spike proteins respectively) 
the area per molecule and $\delta\sim 1\,{\rm nm}$ its typical height fluctuation from the surface.
\end{thebibliography}
\end{document}